\documentclass[manuscript]{aastex}

\usepackage{graphics,color}
\usepackage{ulem}
\definecolor{red}{rgb}{1,0,0}

\shorttitle{Super-Fast-Rotator}
\shortauthors{Chang et al.}
\begin{document}
\title{A New Large Super-Fast Rotator: (335433) 2005 UW163}

\author{Chan-Kao Chang\altaffilmark{1}; Adam Waszczak\altaffilmark{2}; Hsing-Wen Lin\altaffilmark{1}; Wing-Huen Ip\altaffilmark{1,3};
Thomas. A. Prince\altaffilmark{4}; Shrinivas R. Kulkarni\altaffilmark{4}; Russ Laher\altaffilmark{5};
Jason Surace\altaffilmark{5}}

\altaffiltext{1}{Institute of Astronomy, National Central University, Jhongli,Taiwan}
\altaffiltext{2}{Division of Geological and Planetary Sciences, California Institute of Technology,
Pasadena, CA 91125, USA} \altaffiltext{3}{Space Science Institute, Macau University of Science and
Technology, Macau} \altaffiltext{4}{Division of Physics, Mathematics and Astronomy, California
Institute of Technology, Pasadena, CA 91125, USA} \altaffiltext{5}{Spitzer Science Center, California
Institute of Technology, M/S 314-6, Pasadena, CA 91125, USA}

\email{rex@astro.ncu.edu.tw}

\begin{abstract}
Asteroids of size larger than 150 m generally do not have rotation periods smaller than 2.2 hours.
This spin cutoff is believed to be due to the gravitationally bound rubble-pile structures of the
asteroids. Rotation with periods exceeding this critical value will cause asteroid breakup. Up until
now, only one object, 2001 OE84, has been found to be an exception to this spin cutoff. We report the
discovery of a new super-fast rotator, (335433) 2005 UW163, spinning with a period of 1.290 hours and
a lightcurve variation of $r'\sim0.8$ mag from the observations made at the P48 telescope and the P200
telescope of the Palomar Observatory. Its $H_{r'} = 17.69 \pm 0.27$ mag and multi-band colors (i.e.,
$g'-r' = 0.68\pm0.03$ mag, $r'-i' = 0.19\pm0.02$ mag and SDSS $i-z = -0.45$ mag) show it is a V-type
asteroid with a diameter of $0.6 +0.3/-0.2$ km. This indicates (335433) 2005 UW163 is a super-fast
rotator beyond the regime of the small monolithic asteroid.
\end{abstract}


\section{Introduction}
Asteroids with the dwarf planet, Ceres, as the largest member, have sizes ranging from a few meters to
a few hundred km. They were subject to strong collisional interaction, as indicated by the formation
of many asteroid families composed of breakup fragments from disruptive impact events. A fundamental
discovery made by \citet{Harris1996} concerning the internal structures of asteroids has to do with
the statistical result that objects with sizes larger than 150 m have rotation periods longer than 2.2
hours. This spin cutoff can be explained by the rubble-pile structure in which the asteroids are made
up of collisional breakup fragments bound together by mutual gravitational force. Faster rotation will
lead to centrifugal disruption. Only one object, 2001 OE84, which is a near-Earth asteroid with a
diameter of 0.65 km \citep{Warner2009} and shows a rotation period of 29.19 minutes, has been found to
be exception to this rule \citep{Pravec2002}. \citet{Holsapple2007} subsequently proposed that the
binding of such fast rotating asteroids could be the result of size-dependent material strength and
predicted the presence of km/sub-km-sized super-fast rotators (hereafter SFRs). It is noteworthy to
point out that several SFR candidates of this size range were reported by \citet{Masiero2009} and
\citet{Dermawan2011}, but none of them had reliably determined periods \citep{Harris2012}. This is
because their low brightness and fast rotation make such SFRs difficult to be detected and confirmed
by follow-up observations. A comprehensive program is required to produce quantitative measurements.

Within the framework of the PTF/iPTF project\footnote{The Palomar Transient Factory/ intermediate
Palomar Transient Factory; http://ptf.caltech.edu/iptf} and the TANGO project\footnote{Taiwan New
Generation OIR Astronomy}, a systematic survey of asteroid lightcurves has been carried out by data
mining of the archived data \citep[see][for the pilot study]{Polishook2012} and dedicated
observational campaigns \citep[see][for the result for the first campaign]{Chang2014} using the
48-inch Oschin Schmidt telescope (P48). In our second (January 6--9, 2014) and third (February 20--23,
2014) dedicated observational campaigns, a number of SFR candidates have been identified in the
analysis of these data sets (the result including asteroid rotation periods and SFR candidates will be
published in the near future). One of them is (335433) 2005 UW163. In this work, we report on the P48
discovery observation during February 20--23, 2014, and the follow-up observations conducted on the
200-inch Hale Telescope at the Palomar Observatory on March 25, 2014 to confirm the identity of this
interesting object. In Section 2, we describe our observations and data reduction. The rotation period
analysis is given in Section 3. The results and discussion are presented in Section 4. A summary and
conclusion can be found in Section 5.

\section{Observations and Data Reduction}
\subsection{PTF Photometry}
The PTF is a synoptic survey designed to explore the transient and variable sky \citep{Law2009,
Rau2009}. It employs the 48-inch Oschin Schmidt Telescope equipped with an 11-chip mosaic CCD camera
that has a field of view of $\sim7.26$ \,deg$^2$ and a pixel scale of 1.01\arcsec. The available
filters include Mould-{\it R}, Gunn-{\it g\arcmin} and $H_{\alpha}$, and its $5\sigma$ median limiting
magnitude of an exposure of 60 s in $R$-band is $\sim$21 mag \citep{Law2010}. Each PTF exposure was
processed by the IPAC-PTF photometric pipeline including image splitting, de-biasing, flat-fielding,
source extraction, astrometric calibration, and photometric calibration \citep{Grillmair2010}, and
\citep{Laher2014}. Its final products include reduced images, mask images and source catalogs. The PTF
absolute photometric calibration was done by fitting to catalog data in Sloan Digital Sky Survey
fields \citep[][; hereafter, SDSS]{York2000}, which routinely reached a precision of $\sim0.02$ mag
\citep{Ofek2012a, Ofek2012b}. Since the calibration was done on a per-night, per-field, per-filter,
per-chip basis, small photometric zero-point variations are possible from night to night, the accuracy
of which depends on the degree to which a night is affected by mist and cloud cover, and transient
variations in atmospheric conditions.

\begin{figure}
  \plotone{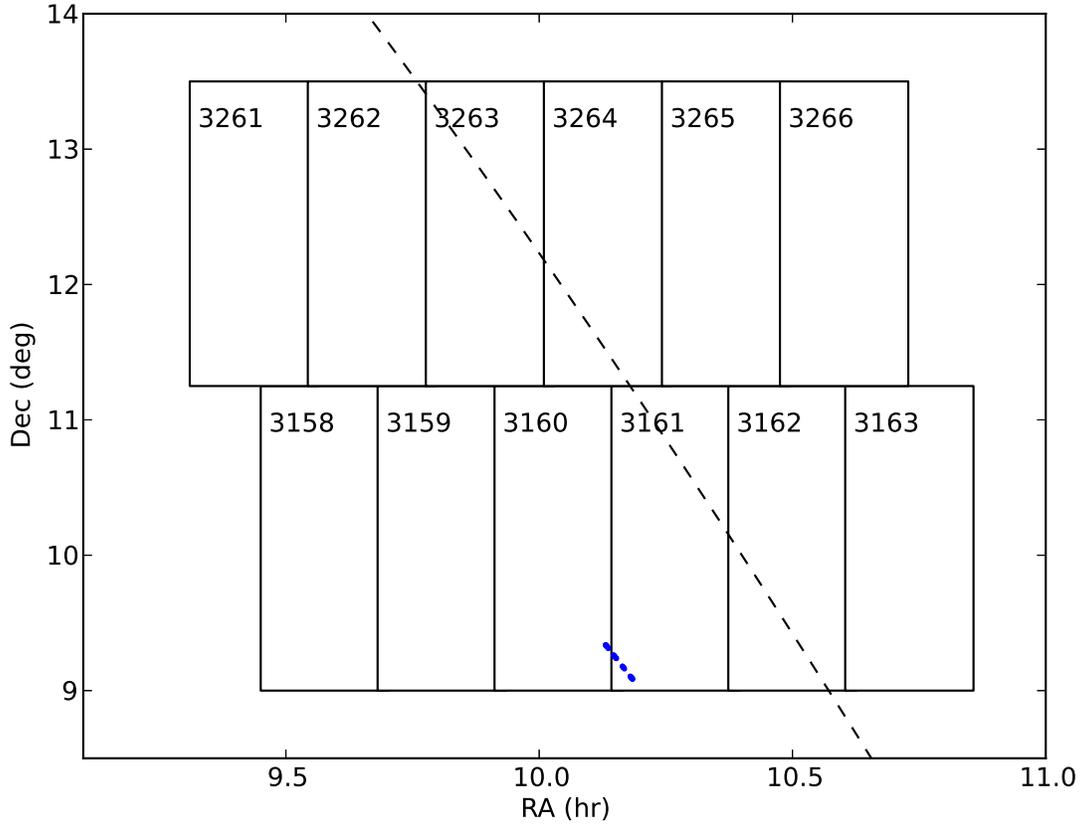}
  \caption{The configuration of 12 PTF fields. Each rectangle represents a PTF field with its field ID.
  The size of each PTF field is $\sim 7.26$ \,deg$^2$. The dashed line
  is the ecliptic plane and the blue dots are the trail of (335433) 2005 UW163.
  Note that the scales of R.A. and Declination are not in proper ratio.}
  \label{obs_fig}
  \end{figure}

As part of the TANGO project, we conducted an asteroid rotation-period survey during February 20--23,
2014, which continuously scanned twelve consecutive PTF fields on the ecliptic plane in $R$-band with
a cadence of $\sim20$ min and an exposure time of 60 s for each frame. The total sky area was $\sim87$
\,deg$^2$. Table~\ref{obs_log_ptf} lists the observational metadata and Fig.~\ref{obs_fig} shows the
field configuration. To extract known asteroid lightcurves, the source catalogs were purged of all
stationary sources and then matched against the ephemerides obtained from the {\it JPL/HORIZONS}
system with a radius of 2\arcsec. In this campaign, we extracted $\sim$2500 lightcurves suitable for
rotation period analysis. The preliminary result showed $\sim$910 objects had good rotation period
determinations, and 11 of them were identified as SFR candidates. The result will be published in a
separate paper soon. The follow-up observations for the SFR candidates were also planned. Our
main-belt asteroid of interest, (335433) 2005 UW163, was close to its opposition and located on the
chips 7,6,6 of field 3161 on 20,21,and 22 Feb 2014, and on chip 11 of field 3160 on 23 Feb 2014.
Therefore, we were able to obtain its lightcurve over a four-night time span to identify it as a
highly probable SFR and conducted a follow-up observation to confirm its short rotation period.

\subsection{The Follow-Up Observation}
The follow-up observation was carried out in the night of March 25, 2014 by using the 200-inch Hale
Telescope at Palomar (hereafter P200), equipped for imaging with the Large Format Camera
\citep[LFC;][]{Simcoe2000}, which has a pixel size of 0.363\arcsec. In order to obtain the lightcurve
and color of (335433) 2005 UW163, we followed an observational filter sequence of 2 $g'$-, 2 $r'$-, 2
$i'$-, 18 $r'$-, 2 $g'$-, 2 $r'$-, 2 $i'$- and 15 $r'$-bands. The time difference between two
exposures was $\sim3$ min, including a 2-min exposure time and an $\sim$1-min readout time. Therefore,
we obtained 47 exposures in total (i.e., 4 $g'$, 4 $i'$ and 39 $r'$ bands) over a time span of
$~\sim150$ min. We applied standard de-biasing, flat-fielding and astrometric calibration to all
images and used SExtractor \citep{Bertin1996} to extract sources. The absolute photometry of each
exposure was calibrated against SDSS point sources of $r' \sim 18$ to 22 mag by using
linear-least-square fitting and typically obtained a fitting residual of $\sim0.02$ mag. Since
(335433) 2005 UW163 would have a trail length of $\sim2\arcsec$ for a 2-min exposure, similar to the
seeing size (i.e., $\sim2\arcsec$) and much larger than the pixel size (i.e., 0.363\arcsec), we
employed the trail-fitting method \citep{Veres2012} to improve its photometric accuracy. The first one
and last two exposures were not included in the following analysis due to their fuzzy images.

\section{Rotation Period Analysis}
To measure the synodic rotation period of (335433) 2005 UW163, all data points were corrected for
light-travel time (i.e., the time interval of a photon traveling from object to Earth) and the
magnitudes were reduced to both heliocentric ($r$) and geocentric $\triangle$ distances at 1 AU by
\begin{equation}\label{reduced_mag}
  M(r=1,\triangle=1) = m + 5\log(r\triangle),
\end{equation}
where $M$ and $m$ are reduced and apparent magnitudes, respectively. The $r$ and $\triangle$ were
calculated by the PyEphem\footnote{http://rhodesmill.org/pyephem/} using the orbital elements obtained
from the Minor Planet Center\footnote{http://minorplanetcenter.net}.

We applied a second-order Fourier series, which was developed by \citet{Harris1989} and has since
become the standard for asteroid lightcurve fitting \citep[e.g.][]{Pravec2005}, to the PTF $R$-band
lightcurve and the P200 $r'$-band lightcurve separately, in order to search for their rotation
periods:
\begin{equation}\label{FTeq}
  M_{i,j} = \sum_{k=1,2}^{N_k} B_k\sin\left[\frac{2\pi k}{P} (t_j-t_0)\right] + C_k\cos\left[\frac{2\pi k}{P} (t_j-t_0)\right] + Z_i,
\end{equation}
where $M_{i,j}$ is the reduced magnitude measured at the light-travel time corrected epoch $t_j$,
$B_k$ and $C_k$ are the Fourier coefficients, $P$ is the rotation period and $t_0$ is an arbitrary
epoch. Following \citet{Polishook2012}, we also fitted a constant value $Z_i$ in Eq.~ (\ref{FTeq}) to
correct the small systematic offsets between different PTF data sets, where a data set was defined as
all the measurements taken on the same night, field, filter and CCD. Then, Eq.~(\ref{FTeq}) was solved
by using least-squares minimization for each given $P$ to obtain the other free parameters. We tried a
frequency range of 0.25--25 rev/day with a step of 0.0025 rev/day to cover the majority of asteroid
rotation periods \citep{Pravec2000}. The uncertainty of the derived rotation period was calculated
from the range of periods with $\chi^2$ smaller than $\chi_{best}^2+\triangle\chi^2$, where
$\chi_{best}^2$ is the $\chi^2$ of the picked-out period and $\triangle\chi^2$ is calculated from the
inverse $\chi^2$ distribution, assuming $1 + 2N_k + N_i$ degrees of freedom.

\section{Results and Discussion}
\begin{figure}
  \plotone{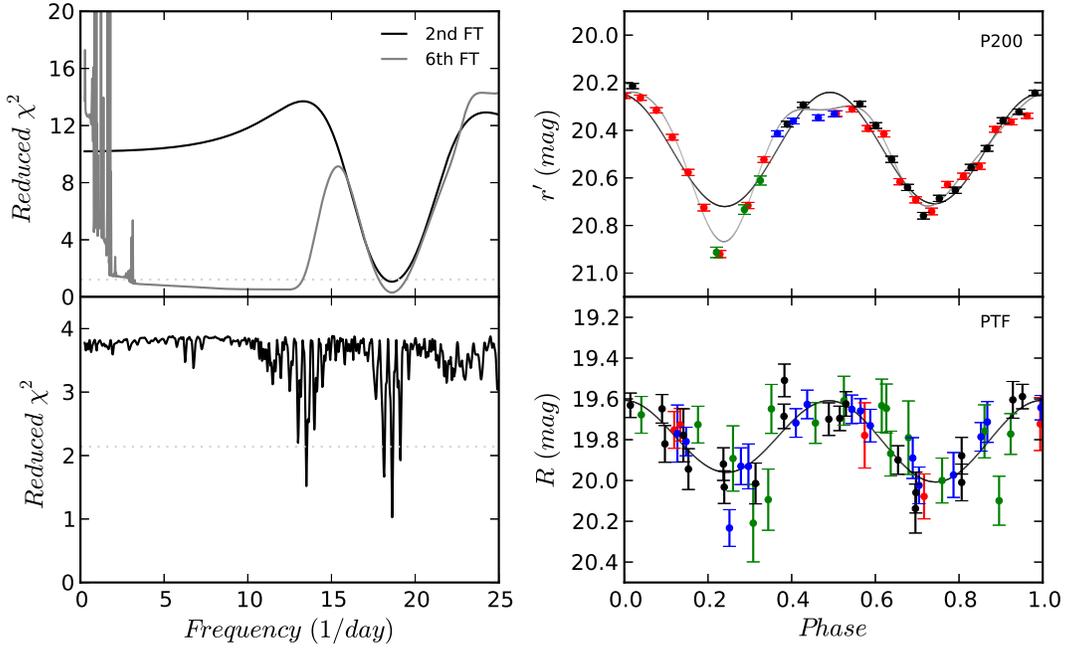}
  \caption{The periodograms (right) and the folded lightcurves (left) of $P = 1.290$ hours (i.e., $f = 18.6$ rev/day)
  for (335433) 2005 UW163 obtained from the P200 (upper) and the PTF (lower) observations. The black and gray lines
  are the results of the second- and  sixth-order Fourier fittings, respectively. The dotted-gray lines in
  the periodograms indicate the uncertainties of the derived rotation periods. The green, blue, red and black filled
  circles in the P200 lightcurve are $g'$, $i'$, the first cycle $r'$ and the second cycle $r'$-band measurements, respectively,
  in which the $g'$-band data are offset by -0.68 mag and that of $i'$-band are offset by 0.19 mag.
  The green, red, blue and black filled circles in the PTF lightcurve are the measurements taken on February 20, 21, 22 and 23, 2014, respectively.}
  \label{lc}
  \end{figure}

The rotation periods of (335433) 2005 UW163 derived from the PTF and the P200 observations are very
compatible with each other, which are $1.290\pm0.034$ and $1.290\pm0.058$ hours, respectively. The
super-fast rotating nature of this asteroid is very convincingly demonstrated in both folded
lightcurves (see Fig.~\ref{lc}). The peak-to-peak variation of the PTF lightcurve is $R\sim0.6$ mag
and that of the P200 lightcurve is $r'\sim0.8$ mag. Although the P200 folded lightcurve does not cover
two full cycles of the derived period, we do not see any particular offset in the overlap between the
first (red filled circles) and second (black filled circles) cycles. Moreover, the lightcurve shows a
large amplitude. Therefore, it is very unlikely that (335433) 2005 UW163 would be the case as
discussed in \citet{Harris2014}, where the authors pointed out that some asteroids might have
ambiguous rotation period determinations due to their small lightcurve amplitudes (i.e., the half
rotation period was misidentified as a ``real'' rotation period). One special case reported by
\citet{Polishook2010} that (40701) 1999 RG235 showed a possible three-peaked lightcurve with a large
amplitude. This could be a possibility for (335433) 2005 UW163. If the lightcurve of our target has a
four-peaks, then it must have identical odd maxima, even maxima and even minima as seen in our case.
Such a chance seems to be vary small, and we thus believe our derived rotation period is highly
probable to be the true rotation period of (335433) 2005 UW163.

In order to improve the P200 lightcurve fitting, we used a sixth-order Fourier series to analysis its
rotation period again. The result is shown in Fig.~\ref{lc} as well. We see the best-fit rotation
period still remains at 1.29 hours and the fine features on the P200 folded lightcurve are reproduced
relatively well as pointed out by \citet{Harris2014}. This means (335433) 2005 UW163 has a more
complex shape.

Since the $g'$ and $i'$ data points show similar variation to the P200 $r'$-band lightcurve, we use
Eq.~(\ref{FTeq}) with the derived rotation period, 1.290 hours, to obtain the offsets, $Z_i$, between
the measurements for the $g'$-, $r'$- and $i'$-bands, where a data set was then defined as the
measurements taken with the same filter. The best-fitting results were $g'-r' = 0.68 \pm 0.03$ mag and
$r'-i' = 0.19 \pm 0.02$ mag, which indicates a S- or V-type asteroid for (335433) 2005 UW163
\citep{Ivezic2001}. Incorporating $i-z = -0.45$ obtained from the SDSS moving object catalog
\citep{Ivezic2002}, we believe (335433) 2005 UW163 is most likely a V-type asteroid
\citep{Parker2008}.



Since the PTF and P200 lightcurves were observed in different bands and the phase angles ($\alpha$)
had a small change in both observations, we only applied, instead of fitting, the $H$--$G$ system
\citep[Eq.~(\ref{Hmag})][]{Bowell1989} with a fixed $G$ slope of $0.43 \pm 0.08$ \citep[i.e., a
typical $G$ value for V-type asteroid;][]{Warner2009} to the P200 $r'$-band lightcurve to estimate the
absolute magnitude $H$ of (335433) 2005 UW163.
\begin{equation}\label{Hmag}
  H = \langle M(r=1,\triangle=1)\rangle+2.5\log[(1-G)\phi_1+G\phi_2],
\end{equation}
where
\begin{equation}
  \phi_1 = \exp [-3.33 \tan(0.5\langle \alpha \rangle)^{0.63}],
\end{equation}
\begin{equation}
  \phi_2 = \exp [-1.87 \tan(0.5\langle \alpha \rangle)^{1.22}].
\end{equation}
This gives a mean value of $H_{r'} = 17.69 \pm 0.27$ mag. Moreover, its diameter is estimated using
\begin{equation}
  D = {1130 \over \sqrt{p_{r'}}} 10^{-H_{r'}/5},
\end{equation}
where $p_{r'}$ is the $r'$-band albedo and the conversion constant, 1130, adopted from
\citet{Jweitt2013}. We assumed an albedo of $p_{r'} = 0.297 \pm 0.131$, which was adopted from the
geometric albedo in visible wavelengths ($p_v$) appropriate for V-type asteroid \citep{Usui2013}.
Although there is a small difference between $p_{r'}$ and $p_v$, we treat this difference as a part of
uncertainty in our assuming albedo. Therefore, we have a derived diameter of $D = 0.6 +0.3/-0.2$ km,
where the diameter range is estimated by the uncertainties introduced by $p_{r'}$, $G$ and $H_{r'}$.

\begin{figure}
  \plotone{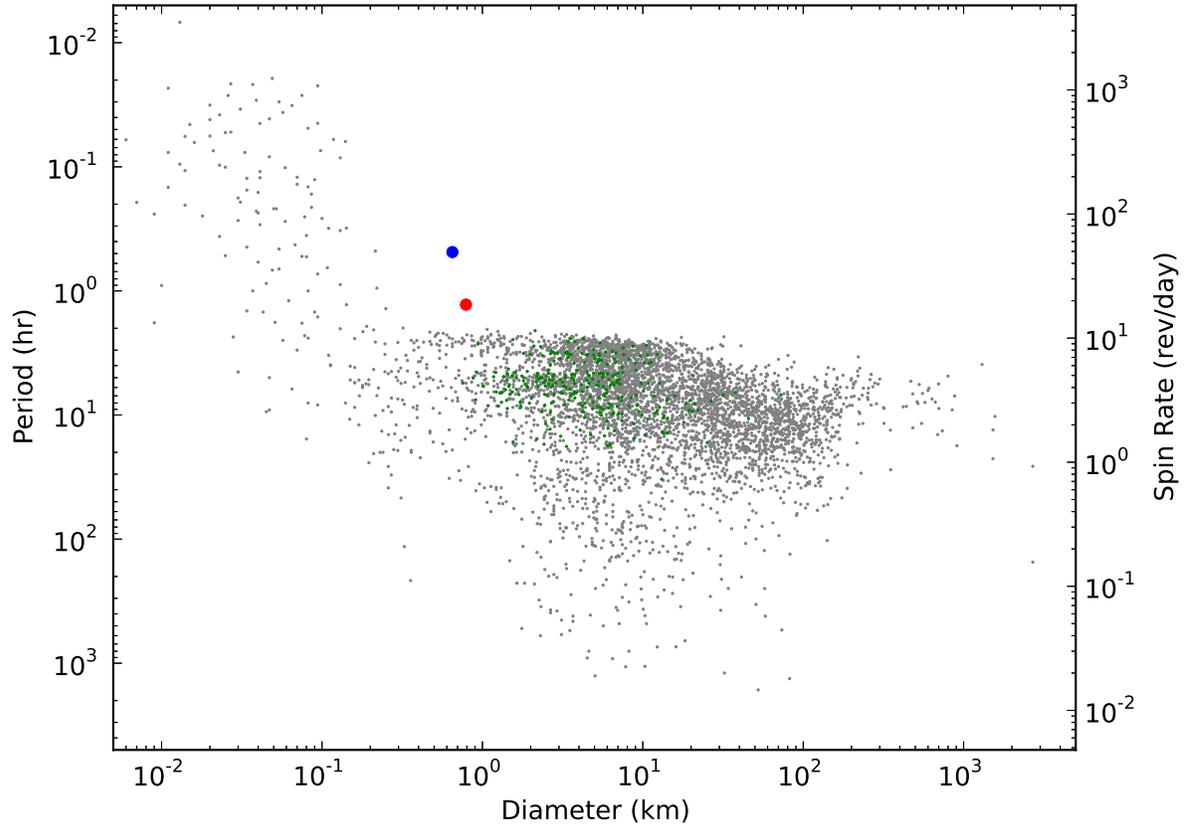}
  \caption{The plot of the diameters vs. rotation period. The green and gray filled circles are the objects with
  good rotation period determinations obtained from \citet{Chang2014} and the LCDB \citep{Warner2009}, respectively.
  The asteroids (335433) 2005 UW163 (red filled circle) and 2001 OE84 (blue filled circle) locate above the ``spin-barrier''
  and away from the small monolithic SFR.}
  \label{dia_per}
  \end{figure}

\begin{figure}
  \plotone{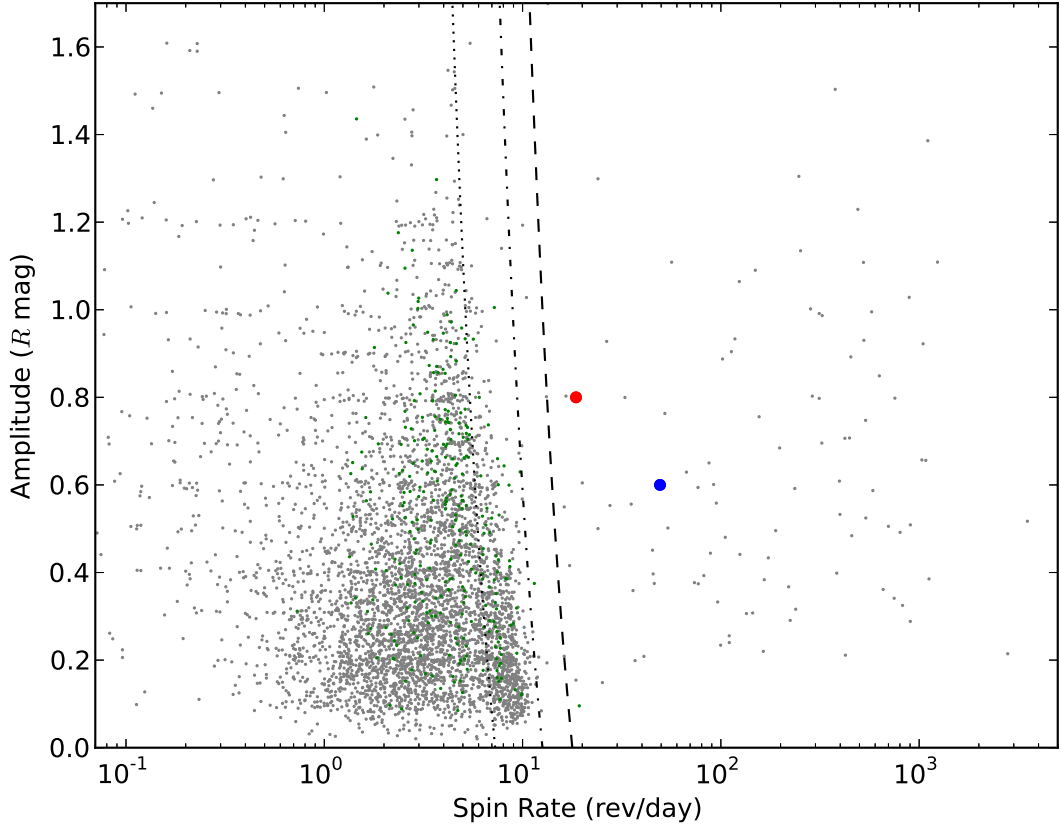}
  \caption{The plot of the spin rate vs. amplitude. The green and gray filled circles are the objects with
  good rotation period determinations obtained from \citet{Chang2014} and the LCDB \citep{Warner2009}, respectively.
  The dashed, dot-dashed and dotted lines represent the spin rate limits estimated by $P \sim 3.3 \sqrt{(1+\Delta m)/\rho}$ hours for ``rubble pile''
  asteroids with bulk densities of $\rho =$ 6, 3 and 1 g/cm$^3$ and lightcurve amplitude, $\Delta m$ \citep{Pravec2000}.
  The asteroids (335433) 2005 UW163 (red filled circle) and 2001 OE84 (blue filled circle) have a bulk density larger
  than 6 g/cm$^3$.}
  \label{spin_amp}
  \end{figure}

Combining the data taken from \citet{Warner2009} and \citet{Chang2014}, we plot the asteroid spin rate
vs. diameter in Fig.~\ref{dia_per}. We see both (335433) 2005 UW163 and 2001 OE84 located at above the
``spin barrier'' of the ``rubble-pile'' asteroids and away from the small monolithic SFRs. This result
indicates (335433) 2005 UW163 is a SFR. To keep (335433) 2005 UW163 from breaking apart under such
fast rotation by its own gravity only, it requires a bulk density as high as $\rho\sim11$ g/cm$^3$
when assuming a critical rotation period as $P \sim 3.3 \sqrt{(1+\Delta m)/\rho}$\,hours for an object
with a lightcurve variation of $r'\sim0.8$ mag \citep{Pravec2000}. Such high bulk density is very
unusual for the ``rubble-pile'' asteroids. This is because, as shown in the plot of spin rate vs.
lightcurve amplitude in Fig.~\ref{spin_amp}, the ``rubble-pile'' asteroids cannot have a bulk density
$> 3$ g/cm$^3$ \citep{Pravec2000}. To explain the presence of km/sub-km-sized SFRs,
\citet{Holsapple2007} used a size-dependent strength for asteroids, which included tensile strength
and cohesiveness, in addition to gravity. This additional size-dependent effect could produce a
transition from the small monolithic object region to the large gravitationally bounded aggregation
region. Therefore,a comprehensive survey of the population of km/sub-km-sized SFRs would help us
understand this question.

\section{Summary and Conclusion}
Sub-km-sized SFR, (335433) 2005 UW163, which is a main-belt asteroid showing a rotation period of
1.290 hours (i.e., a frequency of 18.6 rev/day), was discovered in the asteroid rotation period survey
conducted on the PTF during Feb 20--23, 2014 as part of the TANGO project. It was confirmed by the
follow-up observation carried out on March 25, 2014 using the P200. Its multi-band colors (i.e.,
$g'-r' = 0.68 \pm 0.03$ mag, $r'-i' = 0.19 \pm 0.02$ mag and the SDSS $i-z = -0.45$) indicate that
(335433) 2005 UW163 is a V-type asteroid. With an assumed albedo of $0.297\pm0.131$, the SFR
asteroid's $H_{r'} = 17.69 \pm 0.27$ mag allows us to infer a diameter of $D = 0.6 +0.3/-0.2$ km. This
shows if (335433) 2005 UW163 is a ``rubble pile'' asteroid, it would require a bulk density as high as
11 g/cm$^3$ to keep itself from breaking apart under such fast rotation. Therefore, other mechanisms,
such as a combination of tensile strength and cohesiveness, should be taken into account
\citep{Holsapple2007}. To search for a larger sample of km/sub-km-sized SFRs by comprehensive asteroid
rotation surveys, as is being planned would be very helpful to the investigation of the origin and
internal structures of asteroids.

\acknowledgments This work is supported in part by the National Science Council of Taiwan under the
grants NSC 101-2119-M-008-007-MY3. We are thankful for the indispensable supports provided by the
staff of the Palomar Observatory. We also thank the referee, Alan Harris, for useful comments that
helped to improve the content of the paper.

\begin{deluxetable}{lrrcccc}
\tabletypesize{\scriptsize} \tablecaption{The PTF observation. \label{obs_log_ptf}} \tablewidth{0pt}
\tablehead{ \colhead{Field ID} & \colhead{RA} & \colhead{Dec.} & \colhead{Feb 20} &
\colhead{Feb 21} & \colhead{Feb 22} & \colhead{Feb 23} \\
\colhead{} & \colhead{($^{\circ}$)} & \colhead{($^{\circ}$)} & \colhead{$\Delta$t, N$_\textrm{exp}$} &
\colhead{$\Delta$t, N$_\textrm{exp}$} & \colhead{$\Delta$t, N$_\textrm{exp}$} & \colhead{$\Delta$t,
N$_\textrm{exp}$}} \startdata
    3158 &  143.65  &  10.12 &     7.9, 19 &    7.6, 22 &    7.9, 22 &    7.2, 20 \\
    3159 &  147.12  &  10.12 &     8.0, 17 &    7.3, 20 &    7.9, 21 &    7.7, 22 \\
    3160 &  150.58  &  10.12 &     7.6, 18 &    7.7, 21 &    7.7, 22 &    7.7, 22 \\
    3161 &  154.04  &  10.12 &     8.0, 20 &    8.0, 22 &    7.9, 23 &    7.9, 23 \\
    3162 &  157.50  &  10.12 &     8.3, 20 &    7.9, 22 &    8.0, 24 &    7.9, 23 \\
    3163 &  160.96  &  10.12 &     8.0, 19 &    7.8, 22 &    8.0, 24 &    8.0, 24 \\
    3261 &  141.55  &  12.38 &     7.7, 20 &    7.6, 21 &    8.2, 23 &    7.1, 20 \\
    3262 &  145.05  &  12.38 &     8.1, 20 &    7.9, 23 &    8.2, 23 &    7.5, 20 \\
    3263 &  148.54  &  12.38 &     8.2, 20 &    7.6, 22 &    8.1, 23 &    7.7, 21 \\
    3264 &  152.04  &  12.38 &     7.9, 20 &    8.0, 23 &    8.2, 24 &    8.2, 24 \\
    3265 &  155.53  &  12.38 &     8.7, 19 &    8.0, 22 &    8.1, 24 &    8.2, 23 \\
    3266 &  159.03  &  12.38 &     8.0, 21 &    8.2, 23 &    8.1, 25 &    8.2, 24 \\
\enddata
\tablecomments{$\Delta$t is the time duration spanned by each observing set in hours and
N$_\textrm{exp}$ is the total number of exposures for each night and field.}
\end{deluxetable}


\begin{thebibliography}{}
\bibitem[Bertin \& Arnouts(1996)]{Bertin1996} Bertin, E., \& Arnouts, S.\ 1996, \aaps, 117, 393
\bibitem[Bowell et al.(1989)]{Bowell1989} Bowell, E., Hapke, B., Domingue, D., et al.\ 1989, Asteroids II, 524
\bibitem[Chang et al.(2014)]{Chang2014} Chang, C.-K., Ip, W.-H., Lin, H.-W., et al.\ 2014, \apj, 788, 17
\bibitem[Dermawan et al.(2011)]{Dermawan2011} Dermawan, B., Nakamura, T., \& Yoshida, F.\ 2011, \pasj, 63, 555
\bibitem[Grillmair et al.(2010)]{Grillmair2010} Grillmair, C.~J., Laher, R., Surace, J., et al.\ 2010, Astronomical Data Analysis Software and Systems XIX, 434, 28
\bibitem[Harris et al.(1989)]{Harris1989} Harris, A.~W., Young, J.~W., Bowell, E., et al.\ 1989, \icarus, 77, 171
\bibitem[Harris(1996)]{Harris1996} Harris, A.~W. 1996, Lunar and Planetary Institute Science Conference Abstracts, 27, 493
\bibitem[Harris et al.(2012)]{Harris2012} Harris, A.~W., Pravec, P., \& Warner, B.~D.\ 2012, \icarus, 221, 226
\bibitem[Harris et al.(2014)]{Harris2014} Harris, A.~W., Pravec, P., Gal{\'a}d, A., et al.\ 2014, \icarus, 235, 55
\bibitem[Holsapple(2007)]{Holsapple2007} Holsapple, K.~A.\ 2007, \icarus, 187, 500
\bibitem[Ivezi{'c} et al.(2001)]{Ivezic2001} Ivezi{'c}, {\v Z}., Tabachnik, S., Rafikov, R., et al.\ 2001, \aj, 122, 2749
\bibitem[Ivezic et al.(2002)]{Ivezic2002} Ivezic, Z., Juric, M., Lupton, R.~H., Tabachnik, S., \& Quinn, T.\ 2002, \procspie, 4836, 98
\bibitem[Jewitt et al.(2013)]{Jweitt2013} Jewitt, D., Ishiguro, M., \& Agarwal, J.\ 2013, \apjl, 764, L5
\bibitem[Laher et al.(2014)]{Laher2014} Laher, R.~R., Surace, J., Grillmair, C.~J., et al.\ 2014, arXiv:1404.1953
\bibitem[Law et al.(2009)]{Law2009} Law, N.~M., Kulkarni, S.~R., Dekany, R.~G., et al.\ 2009, \pasp, 121, 1395
\bibitem[Law et al.(2010)]{Law2010} Law, N.~M., Dekany, R.~G., Rahmer, G., et al.\ 2010, \procspie, 7735
\bibitem[Masiero et al.(2009)]{Masiero2009} Masiero, J., Jedicke, R., {\v D}urech, J., et al.\ 2009, \icarus, 204, 145
\bibitem[Ofek et al.(2012a)]{Ofek2012a} Ofek, E.~O., Laher, R., Law, N., et al.\ 2012a, \pasp, 124, 62
\bibitem[Ofek et al.(2012b)]{Ofek2012b} Ofek, E.~O., Laher, R., Surace, J., et al.\ 2012b, \pasp, 124, 854
\bibitem[Parker et al.(2008)]{Parker2008} Parker, A., Ivezi{'c}, {\v Z}., Juri{'c}, M., et al.\ 2008, \icarus, 198, 138
\bibitem[Polishook(2010)]{Polishook2010} Polishook, D.\ 2010, Minor Planet Bulletin, 37, 65
\bibitem[Polishook et al.(2012)]{Polishook2012} Polishook, D., Ofek, E.~O., Waszczak, A., et al.\ 2012, \mnras, 421, 2094
\bibitem[Pravec \& Harris(2000)]{Pravec2000} Pravec, P., \& Harris, A.~W.\ 2000, \icarus, 148, 12
\bibitem[Pravec et al.(2002)]{Pravec2002} Pravec, P., Ku{\v s}nir{'a}k, P., {\v S}arounov{'a}, L., et al.\ 2002, Asteroids, Comets, and Meteors: ACM 2002, 500, 743
\bibitem[Pravec et al.(2005)]{Pravec2005} Pravec, P., Harris, A.~W., Scheirich, P., et al.\ 2005, \icarus, 173, 108
\bibitem[Rau et al.(2009)]{Rau2009} Rau, A., Kulkarni, S.~R., Law, N.~M., et al.\ 2009, \pasp, 121, 1334
\bibitem[Simcoe et al.(2000)]{Simcoe2000} Simcoe, R.~A., Metzger, M.~R., Small, T.~A., \& Araya, G.\ 2000, Bulletin of the American Astronomical Society, 32, 758
\bibitem[Usui et al.(2013)]{Usui2013} Usui, F., Kasuga, T., Hasegawa, S., et al.\ 2013, \apj, 762, 56
\bibitem[Vere{\v s} et al.(2012)]{Veres2012} Vere{\v s}, P., Jedicke, R., Denneau, L., et al.\ 2012, \pasp, 124, 1197
\bibitem[Warner et al.(2009)]{Warner2009} Warner, B.~D., Harris, A.~W., \& Pravec, P.\ 2009, \icarus, 202, 134
\bibitem[York et al.(2000)]{York2000} York, D.~G., Adelman, J., Anderson, J.~E., Jr., et al.\ 2000, \aj, 120, 1579


\end{thebibliography}
\end{document}